# The galactic evolution of the supernova rates.

E. De Donder and D. Vanbeveren

*Astrophysical Institute, Vrije Universiteit Brussel, Pleinlaan 2, 1050 Brussel*

*Abstract*

Supernova rates (hypernova, type II, type Ib/c and type Ia) in a particular galaxy depend on the metallicity (i.e. on the galaxy age), on the physics of star formation and on the binary population. In order to study the time evolution of the galactic supernova rates, we use our chemical evolutionary model that accounts in detail for the evolution of single stars and binaries. In particular, supernovae of type Ia are considered to arise from exploding white dwarfs in interacting binaries and we adopt the two most plausible physical models: the single degenerate model and the double degenerate model. Comparison between theoretical prediction and observations of supernova rates in different types of galaxies allows to put constraints on the population of intermediate mass and massive close binaries.

The temporal evolution of the absolute galactic rates of different types of SNe (including the SN Ia rate) is presented in such a way that the results can be directly implemented into a galactic chemical evolutionary model. Particularly for SNIa's the inclusion of binary evolution leads to results considerably different from those in earlier population synthesis approaches, in which binary evolution was not included in detail.

[1] dvbevere@vub.ac.be
[2] ededonde@vub.ac.be

# 1. Introduction.

Supernovae (SN) are among the most energetic processes in the universe and the number of SN observations in different types of galaxies has increased significantly during the last decade (see Cappellaro, 2001, for a review). A comparison between observations and population number synthesis (PNS)[1] simulations may provide important constraints on the evolutionary scenario's of the SN progenitors, and not in the least on the importance of binaries. De Donder and Vanbeveren (1998, paper I) compared observations with the theoretical predicted type II and Ib/c SN population and concluded that the relative rates are a strong function of the progenitor binary population (see also Kalogera and Belczynski, 2001). Kobayashi et al. (2000) studied the temporal evolution of the galactic SN II and the SN Ia rates. For the latter they considered the two most plausible models: the single degenerate (SD) model (Whelan and Iben, 1973; Hachisu et al. 1999a+b hereafter Ha99) and the double degenerate (DD) model (Iben and Tutukov, 1984; Webbink, 1984). However, Kobayashi et al. (2000) did not separate the SN II's from the SN Ib/c's whereas they adopted a very simple population synthesis model. The latter simplification may introduce a significant uncertainty (see also Yungelson and Livio, 1998, 2000). Furthermore, in their paper, Kobayashi et al. (2000) conclude that, when the DD model applies, in an instantaneous starburst the timescale of occurrence[2] of the majority of the SN Ia's is 0.1-0.3 Gyr. As shown in De Donder and Vanbeveren (2002a) this may not be correct (see also section 4).

Supernovae are among the main drivers of the chemical evolution of a galaxy and therefore it is essential to follow the temporal evolution of their rates as accurate as possible. The studies listed above demonstrate that predicted rates depend critically on the binary population. However, in the chemical evolutionary models (CEMs) that have been used in the past to study the chemical evolution of galaxies, binaries are either ignored or they are treated in an oversimplified way. To illustrate, in many cases the SN Ia rate is included as a free parameter (see Greggio and Renzini, 1983) that is fixed to meet some observational constraints, without modeling in detail the evolution of the progenitor systems. An attempt to discuss the effects of binaries on predicted SN Ia rates and their typical timescales was published by Matteucci and Recchi (2001). However the authors did not follow in detail the evolution of the population of the progenitor SN Ia binary population (i.e. a realistic PNS was not done) and the criticisms formulated by Yungelson and Livio (1998, 2000) also apply here.

---

[1] A PNS code calculates the evolution as function of time of a stellar population consisting of single stars and interacting binaries of all types, with any mass ratio and orbital period, in starburst regions and in regions where star formation is continuous in time. The present state of worldwide PNS research is discussed in the proceedings of the conference 'The influence of binaries on stellar population studies'(2001, ed. D. Vanbeveren, Kluwer Academic Publishers, ISBN 0-7923-7104-6).

[2] The timescale of occurrence of a SN obviously equals the total evolutionary lifetime of the progenitor; we will further use the 'timescale of the SN'.

The scope of the present paper is threefold.

- Theoretically predicted SN II and Ib/c rates depend on the details of massive star evolution, and particularly on the effects of stellar wind on the latter. Since 1998, new stellar wind mass loss rate formalisms were proposed which affect significantly stellar evolution. The SN II and Ib/c results of the present paper are calculated with the new stellar evolutionary models and can be considered as an update of our 1998 paper (paper I).
- We illustrate the temporal evolution of the SN rates after an instantaneous starburst for different metallicities. These calculations show the timescale on which different SN types are formed. Particular attention is given at the timescales of the SN Ia events since they determine the onset of SN Ia iron enrichment of galaxies. The SN Ia timescales are studied when the SD and when the DD scenario applies. Notice that the SD timescales can be compared to those proposed by Matteucci and Recchi (2001).
- To study the SN rates as function of metallicity Z in galaxies, we finally combine our PNS code with a galactic star formation scenario and our CEM of galaxies which includes the chemical enrichment effects by single and binary stars (De Donder and Vanbeveren, 2001, 2002a).

In section 2 we summarise our PNS model. The link with the different SN types is outlined in section 3. We first discuss the typical timescale on which the different SN types occur (section 4) and we illustrate them by following the SN rates in starbursts. Finally, in section 5, we present our simulations of the galactic evolution of the SN rates and compare them to observations of SN rates in different types of galaxies.

**2. The PNS model**

The skeletons of most PNS codes with a realistic fraction of interacting binaries are very similar. The number of studies is very large but a representative set of references are contained in the proceedings of the conference "*The Influence of Binaries on Stellar Population Studies*" (ed. D. Vanbeveren, Kluwer Acad. Pub., Dordrecht, ISBN 0-7923-7104-6). Differences in the PNS results performed by different research teams are due to:

- differences in the evolutionary calculations of single stars and of close binaries that are used to perform PNS
- differences in the mass and angular momentum loss formalism of a non-conservative RLOF, common envelope and spiral-in process
- differences in the physics of the SN explosion and how the SN explosion of one component affects the orbital parameters of a binary
- differences in the criteria used to decide when a binary will merge or not.

A description of our PNS code can be found in Vanbeveren et al. (1998a, b, c), Vanbeveren (2000, 2001). Our code was made in the first place to study massive star populations. However, it was straightforward to adapt the code so that it can handle intermediate mass single stars and binaries as well. In its present state it allows to explore a population of binaries with initial parameters $3 < M_1/M_o < 120$ ($M_1$ = primary mass), $0 < q < 1$ (q = binary mass ratio = mass secondary/mass primary) and $P_{min}^3 < P < 10$ years (P = binary orbital period), and a population of single stars with $0.1 < M/M_o < 120$. It accounts for the Roche lobe overflow (RLOF), mass transfer and mass accretion in case A and case $B_r$ binaries, the common envelope (CE) process and spiral-in in case $B_c$ and case C binaries, the CE and spiral-in in binaries with a compact companion (a white dwarf = WD, a neutron star = NS or a black hole = BH) and the effects of an asymmetric SN explosion on the binary parameters when one of the binary components explode. We use a detailed set of stellar evolutionary calculations for different initial metallicities ($0.001 < Z < 0.02$) where, in the case of massive stars, the most recent stellar wind mass loss rate formalisms are implemented (see appendix A). All evolutionary calculations rely on the core overshooting formalism proposed by the Geneva group (Schaller et al., 1992).

---
[3] $P_{min}$ is the minimum initial period that a binary can have; it is calculated by assuming that the primary fills its Roche lobe already on the ZAMS. For most of the binaries, $P_{min} \approx 1$ day.

## 2.1. Initial distributions

- The single star and binary primary initial mass functions (IMF) are assumed to be equal and constant in space and time; they obey the following law: $\varphi(M) \propto M^{-\gamma}$ with $\gamma = 2.3$ (Salpeter, 1955) for $M < 2\ M_o$ and $\gamma = 2.7$ (Scalo, 1986) for $M > 2\ M_o$.
- We use different binary mass ratio distributions $\phi(q)$: a Hogeveen type distribution (Hogeveen, 1991), a flat distribution and one that peaks at $q = 1$ (Garmany et al., 1980). Notice that a Hogeveen type distribution peaks at small q values, which is also predicted by a model where the distribution of the sum of the binary component masses corresponds to the overall IMF.
- The initial binary period distribution $\Pi(P)$ is flat in the logP (Popova et al., 1982; Abt, 1983) between $P_{min}$ and 10 years. Most of the binaries with a period larger than 10 years will not interact and their components are treated as single stars.
- We account for the effects of asymmetric SN explosions on the binary parameters. Due to the latter, a compact SN remnant receives a kick. We assume that the direction of the SN kicks is isotropic and that they have magnitudes that follow a $\chi^2$-like distribution $f(v_{kick})$ (see paper I), corresponding with the observed space velocity distribution of single pulsars from Lorimer et al. (1997). Accounting for possible observational uncertainties, we calculate the survival probability of the binary for two average kick velocities: $<v_{kick}> = 150$ km/s and 450 km/s. i.e.

$$f(v_{kick}) = 1.96 \cdot 10^{-6} v_{kick}^{3/2} e^{-v_{kick}/171}$$

and  (1)

$$f(v_{kick}) = 2.7 \cdot 10^{-5} v_{kick}^{3/2} e^{-v_{kick}/60}$$

- When a BH forms with a preceding SN (see also section 3.1), the kick attributed to the proto-NS is weighted with the amount of fall back material which is equal to the difference in the final BH mass and the mass of the pre-SN iron core.

**2.2. Evolutionary related PNS parameters**

The orbital period evolution of a binary is governed by SW mass loss in the case of massive binaries and by the different binary interaction processes. The following parameters are important:

- $\beta$, the fraction of matter that is lost by the primary star during RLOF and is accreted by the secondary star. We use the following prescription: in case A/B$_r$ systems with q > 0.4, $0 < \beta = \beta_{max} <= 1$. Spiral-in for q < 0.2 binaries implies $\beta = 0$ for these systems and between q = 0.2 and q = 0.4 a linear relation is adopted. The latter is more than sufficient for the present paper.
- When matter leaves a case A/B$_r$ binary during RLOF, the easiest way to do so is via the second Langrangian point L$_2$ to form a ring around the binary. A relation for the resulting period variation is discussed in appendix B.
- $\alpha$, the efficiency of the conversion of orbital energy into potential energy during CE and spiral-in in the formalism of Webbink (1984) adapted by de Kool (1990). Because the energy-efficiency may depend on the structure (evolutionary status) of the components, we distinguish three different CE phases: the CE phase of non evolved binaries with an initial mass ratio q < 0.2 ($\alpha_1$), the CE phase of case Bc/C binaries ($\alpha_2$) and the CE/spiral-in phase of binaries with a compact companion ($\alpha_3$).

## 2.3. The binary formation rate $f_b$

We define the parameter $f_b$ as the formation rate of binaries with the properties given in subsection 2.1 (which corresponds in a star formation model to the fraction of binaries on the zero age main sequence). Notice that most of these binaries will interact, i.e. the primaries in most of these binaries will fill their Roche volume at a certain moment during their evolution.

From observational studies on spectroscopic binaries in the solar neighbourhood we know that about 33% (±13%) of the O-type stars are the primary of a massive close binary with a mass ratio $q > 0.2$ and a period $P < 100$ days (Garmany et al., 1980). A similar conclusion holds for the intermediate mass B-type stars (Vanbeveren et al., 1998). Accounting for observational selection, it can be shown by binary population synthesis studies that to meet the above observations, an initial OB-type binary fraction $f_b$ larger than 50-70% is required (Vanbeveren et al., 1997; Mason et al., 2001, Van Rensbergen, 2001, and references therein).

In section 5 we will discuss the evolution of the SN rates in the solar neighbourhood. Anticipating, to explain the observed SN Ia rate, the overall interacting binary formation rate in the intermediate mass range must be at least 40-50 %. We like to remind that in general, the binary formation rate differs from the observed overall binary fraction in a stellar population. A stellar population consists of evolved and non-evolved stars. An evolved star that is observed as a single star, can be a merged binary or it could have been a secondary of an interacting binary which was disrupted due to the SN explosion of the primary. This means that the (observed) binary fraction in a stellar population is always smaller than the real (past) binary formation rate.

## 3. SN progenitor models

### 3.1. Black hole formation and hypernova

The observed masses of BHs in low mass X-ray binaries (LMXBs) listed by Bailyn et al., (1998) and the large BH mass in the high mass X-ray binary (HMXB) Cyg X-1 provided a main reason to prefer (already in 1998) evolutionary calculations of massive stars with reduced stellar winds during the Wolf-Rayet (WR) phase (see appendix A for the references which contain the results of these early calculations, for a summary and an update/addendum). We concluded that large(r) mass BH (mass larger than 4-5 $M_o$, up to 10 $M_o$ and even larger) are formed from progenitors with an initial mass > 40 $M_o$. Lower mass BHs (mass between 2 $M_o$ and 4-5 $M_o$) are descendants from massive single stars with an initial mass between 25 $M_o$ and 40 $M_o$.

Some of the LMXB-BH candidates and Cyg X-1 have large space velocities which may be an indication that SN like mass ejection occurred prior to BH formation (Brandt et al., 1995; Nelemans et al., 1999; Fryer and Kalogera, 2001). The $\alpha$-elements in the atmosphere of the optical companion star of the LMXB GRO J1655-40 (Nova Sco 1994) observed by Israelian et al., (1999) strongly support the scenario where the BH formation was preceded by some SN-like mass ejection.

Presently three SNe are known where the explanation of the light curve requires an exceptionally large explosion energy SN1998bw, SN1997ef and SN1997cy (Galama et al., 1998, 1999; Bloom et al., 1999; Reichart et al., 1999; van Paradijs, 1999). They are classified as 'hypernovae' (HN) (for a review, see Nomoto, 2001). Hypernovae may be associated with BH formation (Paczynski, 1998; MacFadyen and Woosley, 1999), however, whether or not the majority of the BH's were preceded by a hypernova is questionable. To explain the lightcurve of hypernovae, 0.4-0.7 $M_o$ of $^{56}$Ni is required (Nakamura et al., 2001) but the optical star of GRO J1655-40 contains very little $^{56}$Fe which may indicate that only a small amount of $^{56}$Ni was present in the ejected mass that was accreted by the companion. Furthermore, De Donder and Vanbeveren (2002b) calculated the early evolution of the solar neighbourhood (Z < 0.001) using a chemical evolutionary model that accounts in detail for the effects of binaries (De Donder and Vanbeveren, 2001, 2002a) and assuming that all stars with an initial mass > 40 $M_o$ eject 0.4-0.7 $M_o$ of Fe. Compared to observations, the Fe increment occurs too fast and the [O/Fe] behaviour is not reproduced. From these results we concluded that not all stars with initial mass > 40 $M_o$ end their life as a BH preceded by a hypernova.

For the scope of the present paper, we separately consider the stars that end their life as a BH (i.e. all single stars with initial mass > 25 $M_o$ and all interacting binary components with initial mass > 40 $M_o$, see also De Donder and Vanbeveren, 2001, 2002a) and calculate the hypernova rate as if all BH are preceded by a hypernova. However, accounting for the discussion above, our theoretical rates are most probably generous upper limits.

**3.2. SN II and SN Ib/c**

All massive stars with an initial mass smaller than the corresponding BH limit that have lost their hydrogen rich layers are counted as SN Ib/c progenitors. This includes all primaries with an initial mass larger than 10 $M_o$ of interacting close binaries that avoid merging. The massive stars that end their live with a hydrogen rich envelope are assumed to explode as an SN II. Obviously the predicted rates depend on the stellar wind mass loss rates and on the effect of metallicity on the latter (appendix A).

**3.3. SN Ia and a particular class of SN II**

Since our PNS code is able to calculate in detail the evolution of a population of intermediate mass binaries, it gives us in a straightforward way the population of binaries consisting of a CO-WD and a main sequence (MS) star (who's evolution may have been affected by mass accretion during the RLOF of the mass loser when it was a member of case A/case $B_r$ binary). During its evolution the secondary star may fill its Roche lobe and transfer mass to the WD. When the WD accretes matter mainly three things can happen depending on the accretion rate.

i. For a very restricted range of the accretion rates stable burning on the WD surface can happen and the mass of the WD will increase. These systems have been observationally identified as Super Soft Binary X-ray Sources (van den Heuvel et al., 1992; Kahabka and van den Heuvel, 1997). Depending on its mass prior to accretion the WD may reach the Chandrasekhar mass and explode, possibly as a SNIa in which large amounts of Fe are ejected. This SN Ia model is known as the single degenerate (SD) model. The WD + MS/RG progenitor criteria for which such a scenario can happen have been derived by Ha99. In the derivation of the conditions Ha99 used the 'strong wind' model (Hachisu et al., 1996). In this model accretion onto the WD is regulated by a fast wind that develops when accretion rates of the order of ($10^{-7}$ to $10^{-6}$) $M_o$/yr are encountered. The WD wind increases the binary mass ratio and period interval for which a SNIa can happen. However, the wind is mainly radiation-driven in the iron lines which means that at low metallicity the wind becomes very weak. This metallicity effect has been investigated by Kobayashi et al. (1998) who showed that at $Z < 0.004$ the WD wind dies out implying a strong reduction in the formation rate of SNIa's at low metallicity. In order to investigate the effects of this Z-dependent mass loss rate model we will consider the SD scenario where the WD wind is either Z dependent (we use the model of Ha99) or Z independent (for every Z, we apply the $Z = 0.02$ results of Ha99).

The SD conditions discussed above are implemented into our PNS code so that we can derive in a straightforward way the realisation frequencies of SN Ia formed through the SD model.

ii. When the mass of the MS/RG star is smaller than the masses in the SD model discussed above, the accretion rates will be lower. At low accretion rates the WD undergoes nova explosions whereby some of the inner core is ejected leading to a reduction of the WD mass. They are not important for the scope of the present paper.

iii. When the mass of the MS/RG star is larger than required for the SD model, the accretion rates are larger than permitted. In this case a red-giant-like envelope forms around the WD with stable burning of hydrogen in a shell around the WD. As both stars are now larger than their Roche lobes, the binary enters a stage of CE evolution. The CE will be removed at the expense of orbital energy (Webbink, 1984; de Kool, 1990) and spiral-in starts. The outcome of the spiral-in process is either the coalescence of the WD with the CHeB companion star or a short period WD + hydrogen deficient CHeB star system where, depending on its mass, the CHeB star evolves into a SN or a WD.

The result of the merging process is not known. It was suggested by Sparks & Stecher (1974) on basis of energy computations that a NS may form with a possible SN explosion. Another possibility may be that the intruding CO WD ignites carbon with a flash and completely desintegrates. Whatever the outcome is, if a SN explosion takes place with or without leaving a remnant, it will show hydrogen in its spectrum if during spiral-in the envelope of the companion giant star was not completely ejected. In our PNS code we assume that a SN takes place if the combined mass of the WD and the core of the companion is larger than 1.4 $M_o$. We classify the SN as a spectral type II with the notation SNIIwd.

Of particular importance are the double CO-WD binaries. Their further evolution is driven by the loss of angular momentum and energy via gravitational wave radiation (GWR) which leads to the coalescence of both stars. This merger may produce a SNIa as well if the sum of the masses of both WDs > 1.4 $M_o$. This model is generally referred to as the double degenerate (DD) scenario (Iben and Tutukov, 1984; Webbink, 1984). A major argument against the DD model (for SN Ia) is the lack of observed double WD systems where the sum of the masses > 1.4 $M_o$. However, this argument has been seriously weakened by Yungelson et al. (2001). Another major argument against it is that the merger of two WDs with a combined mass > 1.4 $M_o$ may very well produce a NS in stead of an SNIa (e.g. Segretain et al., 1997).

Our PNS code computes the double WD population. A double WD binary is accounted for as an SNIa producer when the sum of the WD masses > 1.4 $M_o$ and when the system merges within Hubble time (= 15 Gyr). The lifetime of the SN Ia is the sum of the evolutionary lifetime of the system at the moment of formation of the second WD and the characteristic timescale for merging by GWR which is given by

$$\tau_{GWR} = 8 \cdot 10^7 \, (\text{yrs}) \cdot \left( \frac{(M_1 + M_2)^{\frac{1}{3}}}{M_1 \cdot M_2} \right) \cdot P^{\frac{8}{3}} \, (\text{hr}) \qquad (1)$$

(Landau & Lifshitz, 1959).

## 4. The evolution of the SN rates in starbursts

In order to illustrate typical SN timescales, figure 1 shows the temporal evolution of the number of core collapse SNe (= hypernova, SNII and SN Ib/c) for the case of an instantaneous starburst with $\beta_{max} = 1$, $\alpha_1 = \alpha_2 = \alpha_3 = 1$, $<v_{kick}> = 450$ km/s. Figure 1a holds for $Z = 0.02$ whereas Figure 1b shows the results for $Z = 0.002$ starbursts. The plots labelled with 's' hold for a starburst with 100% single stars, whereas the plots labelled with 'b' hold for starbursts with 100% binaries. Since the number of single and binary SNe scale linearly with the single star frequency and the corresponding binary frequency, it is straightforward to predict the SN numbers for a starburst with any single/binary frequency. All distribution functions are normalised in such a way that the total mass of the stars in the starburst equals 1 $M_o$. This choice allows to easily scale our results to any starburst of arbitrarily total stellar mass.

We explored the consequences on the simulation when the values of the PNS parameters differ from those given above. It can be concluded that the results hardly depend on the $\alpha$ values, $\beta_{max}$ and $<v_{kick}>$. Due to the effect of metallicity on the stellar wind mass loss of a massive star, the smaller Z the shorter the I(s) phase and for $Z = 0.002$ it does not even occur. The II(s) and II(b) phase start somewhat earlier when Z is smaller, however, the overall results are very similar as for a starburst with $Z = 0.02$.

A major conclusion is that the core collapse SN phase of a starburst lasts much longer when binaries are present, i.e. 40 Myr with single stars only compared to 250 Myr with binaries. This is mainly due to two processes: the mass transfer in interacting binaries and the coalescence of a WD with its MS/RG companion.

Figure 2 shows the evolution of the SN Ia number for a starburst that consists of 100% binaries (also the SN Ia number scales linearly with the adopted binary frequency so that it is straightforward to estimate the numbers with any binary frequency) and this for the SD and DD model. We only consider the parameter sets 1 and 4 to illustrate the effect of the parameters $\beta_{max}$ and $\alpha$, which are the parameters that most affect the orbital period evolution of the progenitor binary systems. The mass-ratio distribution critically affects the number but not the timescale on which the SNIa's explode. At $Z = 0.002$ we assume a Z-dependent WD wind for the SD scenario.

| set | $\phi(q)$ | $\beta_{max}$ | $\alpha_1$ | $\alpha_2$ | $\alpha_3$ |
|-----|-----------|---------------|------------|------------|------------|
| 1 | F | 1 | 1 | 1 | 1 |
| 2 | G | 1 | 1 | 1 | 1 |
| 3 | H | 1 | 1 | 1 | 1 |
| 4 | H | 0.5 | 0.5 | 0.5 | 0.5 |
| 5 | H | 1 | 0.5 | 0.5 | 1 |

**Table 1**: Different sets of PNS parameters for which computations are made. The label F stands for a flat mass ratio distribution, G for a Garmany et al. (1980) distribution and H for a Hogeveen (1991, 1992) distribution.

*Conclusions*

- Only when we accept the DD scenario as SN Ia progenitor model, the first SN Ia occurs when the starburst is 40 Myr old. With the SD scenario, the first SN Ia happens when the starburst is at least 300 Myr old.

- With the SD model, most of the progenitor binaries went through a CE phase during the first RLOF phase; with the DD model however, a significant number of the progenitors evolved via a (quasi)-conservative first RLOF but went through a CE phase during the second RLOF phase.

- The predicted numbers depend critically on the details of the RLOF phase in case A/B$_r$ binaries (the parameter $\beta_{max}$) and on the details of the common envelope phases in case B$_c$/C systems (the parameter $\alpha$). In the case of the SD scenario, the SN Ia number depends significantly on the primordial metallicity of the starburst.

- When the DD scenario applies, the lifetime of a SN Ia ranges between 40 Myr and > 10 Gyr. Depending on the PNS parameters, the SN Ia number of one stellar population reaches two maxima: one at ~0.3 Gyr and one at ~5 Gyr. The first maximum contains mainly the SN Ia's where the progenitor binaries experienced two CE phases (MS+MS => (*CE*) => WD+MS => (*CE*) => WD+WD). This maximum corresponds with the results obtained by Iben and Tutukov (1984). The second maximum contains the SN Ia's where the progenitor binaries first evolved through a quasi-conservative RLOF, i.e. MS+MS => (*RLOF*) => WD+MS => (*CE*) => WD+WD. They have longer final periods and therefore longer lifetimes. This possibility was not considered by Iben and Tutukov (1984), however, as illustrated by our simulations, it may be an important one. As illustrated by set 4 which corresponds with $\beta_{max}=0.5$, this second group of long period systems is strongly reduced (they are either shifted towards smaller periods or merged) because the system looses a significant amount of orbital momentum when matter leaves the binary during RLOF.

## 5. The evolution of the SN rates in galaxies

Supernova rates in galaxies depend on the physics of galaxy formation, on the overall star formation rate and on stellar evolution. At least the latter depends on the metallicity. Therefore, to calculate the temporal evolution of the SN rates in galaxies, it is essential to combine a star formation model (SFM), a galactic chemical evolutionary model (CEM) and a PNS model (notice that in general, a CEM includes a SFM but since the supernova rates depend critically on the SFM we will consider it separately). The SN rates depend on the binary population and therefore, to be consistent, also the CEM has to account for the evolution of binaries and their chemical yields. The Brussels CEM that accounts for the evolution of binaries has been described in De Donder and Vanbeveren (2001, 2002a). Our CEM uses the galaxy and star formation model of Chiappini et al. (1997) [see also Talbot and Arnett (1975) and Chiosi (1980)]. We explored the effects of binaries on the overall SFM and concluded that, although interacting binaries return less matter to the interstellar medium (due to a higher formation rate of NSs and BHs), the effects of binaries is very small even for a constant binary frequency of 70%. The SFR in the solar neighbourhood as predicted by our CEM is given in figure 3 and should be typical for all spiral galaxies which forms in two phases of major infall discussed in the papers cited above.

Obviously, the absolute SN rates predicted by our simulations depend on the parameters of the models of star formation and galaxy formation. However, relative rates hardly depend on them.

Table 2 gives the relative rates in the galaxy after 15 Gyr (corresponding to the present age of our galaxy) for the different PNS parameters sets (defined in table 1) and figure 4 illustrates their temporal evolution. For the SNIa's we separately give the predictions obtained with the SD model (first number in the two last columns) and the DD model (second number in the two last columns). In case of the SD model we assumed that the WD wind is metallicity dependent. We will compare our theoretical estimations with observations of spiral galaxies that we take from the compilation of Cappellaro et al. (1999). They are summarised in table 3.

| Mod | $f_b$(%) | set | IIwd/II | II/Ibc | HN/(II+Ibc) | Ia/Ibc | Ia/(II+Ibc) |
|---|---|---|---|---|---|---|---|
| 1 | 40 | 1 | 0.12 | 3.81 | 0.147 | 0.24 / 0.82 | 0.05 / 0.17 |
| 2 | 40 | 2 | 0.14 | 3.31 | 0.148 | 0.17 / 0.69 | 0.04 / 0.16 |
| 3 | 40 | 3 | 0.07 | 5.57 | 0.147 | 0.53 / 0.38 | 0.08 / 0.06 |
| 4 | 40 | 4 | 0.05 | 5.98 | 0.149 | 0.21 / 0.05 | 0.03 / 0.01 |
| 5 | 40 | 5 | 0.06 | 6.13 | 0.149 | 0.21 / 0.36 | 0.03 / 0.05 |
| 6 | 70 | 1 | 0.25 | 2.44 | 0.141 | 0.30 / 1.05 | 0.09 / 0.31 |
| 7 | 70 | 2 | 0.27 | 2.03 | 0.142 | 0.24 / 0.90 | 0.08 / 0.30 |
| 8 | 70 | 3 | 0.10 | 4.20 | 0.140 | 0.78 / 0.60 | 0.15 / 0.12 |
| 9 | 70 | 4 | 0.10 | 4.72 | 0.140 | 0.34 / 0.07 | 0.06 / 0.01 |
| 10 | 70 | 5 | 0.11 | 4.92 | 0.140 | 0.35 / 0.65 | 0.06 / 0.11 |

**Table 2:** The relative SN rates in the galaxy after 15 Gyr for the different PNS parameter sets earlier defined in table 1. In the two last columns the first given number is computed with the SD model and the second one with the DD model.

| Galaxy type | Ia/Ibc | II/Ibc |
|---|---|---|
| Early Spirals | 1.6 | 3.75 |
| Late Spirals | 1.5 | 6.12 |

**Table 3:** The local SN type ratios

Assuming that our galaxy is an intermediate type spiral, it follows from table 3 that the ratios are Ia/Ibc = 1.6 and II/Ibc = 4.9. Interestingly, these numbers correspond fairly well with number counts of historical supernovae in the solar neighbourhood (Strom, 1994).

Our simulations and the comparison of the latter with the observations allow to propose the following conclusions.

*Conclusions*

- Depending on the binary properties, between 5% and 30% of all type II SNe may be the result of the coalescence of a WD with a MS/RG star.
- The expected relative rate of hypernovae to core-collapse SNe is about 15% which corresponds in our galaxy with an absolute rate of 2-4 hypernovae per millennium.
- The Ia/(II+Ibc) number ratio increases almost linearly as function of time, independent from the adopted PNS parameters.
- The predicted II/Ib SN rate ratio depends significantly on the properties of the massive binary population (the massive binary frequency and the mass ratio distribution), confirming the results of paper I. However, the ratio hardly depends on the metallicity and, therefore, the ratio in a galaxy varies very slowly as a function of time.
- Since most of the SN Ib progenitors are binary components, the observed ratio Ia/Ibc may be an indication that the massive binary population relative to the intermediate mass binary population is similar in early and late spirals. However the fact that the ratio II/Ib is significantly different could be an indication that the overall binary frequency (or the overall binary population) in both types of spirals is significantly different.
- When we account for all observations of OB type stars in the solar neighbourhood (including the OB type binary population), accounting for statistical biases when interpreting binary data (Hogeveen, 1991,1992; Mason et al., 2001; Van Rensbergen, 2001) which implies that the interacting binary frequency is at least 50% (section 2.3) whereas the binary mass ratio distribution is flat or peaks at small values of q, then a best correspondence is achieved between the simulated and the observed SN population in the solar neighbourhood if:

  - the CE phase is very efficient ($\alpha$ close to 1)
  - the RLOF in intermediate mass binaries with a mass ratio q > 0.4 is almost purely conservative ($\beta_{max} = 1$)

- the coalescence of a WD with the helium/carbon-oxygen core of its companion star produces a supernova of spectral type II,
- both the SD and DD scenario produce SN Ia's.

**7. Comparison to other studies**

Tutukov et al. (1992) computed the SN rates of different spectral types assuming a constant star formation rate over the lifetime of our galaxy and a 100% initial binary frequency. For the SNIa's they only consider the DD scenario. Scaling their results to a binary fraction of 70% we find that their predicted relative rates are close to our values except for the ratio Ia/Ibc for which we find a higher value. The explanation for this difference is twofold.

1) Tutukov et al. (1992) predict a SNIbc rate of (0.6 – 1) per century which is higher than our estimated rate of (0.3 – 0.7) per century. Since they assume a symmetric SN explosion, the majority of their binary systems remains bound after the explosion of the primary massive star. This implies that the secondary star (if massive) in many cases loses its hydrogen envelope via spiral-in of the compact star and explodes as an SNIbc. It is obvious that when the SN explosion occurs asymmetrical (which is assumed in our PNS model) most binary systems (about 80% when using an average kick magnitude of 450 km/s) are disrupted after the explosion of the primary star and the secondary star further evolves as a single star. According to single star evolutionary scenario adopted in our PNS code the secondary star explodes as an SNIbc only if its mass is larger than (17-20) $M_o$ for Z=0.02. This explains the lower SNIbc rate in the present study.

2) In our DD model we predict that a significant fraction of the SNIa's form with lifetimes > 1 Gyr coming from binary progenitors that evolved through a first quasi-conservative RLOF phase and a CE phase during the second RLOF. In the evolutionary scenario of Iben and Tutukov (1984) (which is used in the work of Tutukov et al. (1992)) SNIa's are only produced by binary progenitors that underwent CE evolution during the first and second RLOF. Therefore the fraction of SNIa's formed in their evolutionary scenario with a lifetime > 1 Gyr is small. By consequence our predicted SNIa rate is generally larger up to a factor 1.5-2, depending on the adopted PNS parameter values. Only when using PNS model 4 we find a good correspondence with the results of Tutukov et al.

(1992). In this model $\beta_{max}$=0.5 which strongly reduces the fraction of SNIa's with lifetimes > 1 Gyr as we explained earlier in section 4. We notice that due to our higher predicted rate of SNIa's and lower rate of SNIbc's we better approach the observed Ia/Ibc ratio.

In their PNS study of binary compact objects Belczynski et al. (2002) report II/Ibc ratios between 1.8-3.7 for Z=0.02 and an initial binary fraction of 50%, which agrees with our predicted ratios.

The time evolution of the SNII and SNIa rate has been computed by Kobayashi et al. (2000) for elliptical and spiral galaxies. They propose an explanation for the higher SNII rate (a factor of 2) observed in late-types spirals compared to early-types based on a different star formation rate. However in their galactic evolutionary model they adopt a very simple population synthesis model and they do not separate the SN II's from the SN Ib/c's. The latter assumption makes that their explanation may not be true. A different star formation rate would also imply a different SNIbc rate in both spiral types which is not observed. On this point we think that our proposition of a different overall binary formation rate in both types of spiral galaxies (see previous section) offers a better explanation.

The evolutionary behaviour of the Ia/II ratio (to be compared with our Ia/(II+Ibc) ratio) as a function of time predicted by Kobayashi et al. (2000) is similar to our results for both the DD and SD model, though their predicted ratio-values are generally higher especially during the early evolutionary phases of the galaxy. For the present (t=15 Gyr) ratio we find a best correspondence with their predictions when considering our models that assume a constant binary frequency of 70%. We remark that Kobayashi et al. (2000) uses the evolutionary model of Iben and Tutukov (1984) for the DD scenario. The lack a of a detailed discussion on the adopted stellar evolutionary models in the work of Kobayashi et al. (2000) does not allow for a thorough discussion on the differences between their work and ours.

## 8. A recipe to implement absolute SNIa rates into a CEM

Figure 5 shows the time-evolution of the absolute SN Ia rates in the galaxy for the SD and the DD scenario computed with a constant binary frequency of 70%. The results depend in a linear way on this frequency and on the constants in the SFR model (remember that the temporal evolution of the SFR rate hardly depends on the adopted binary frequency, section 5). This means that the graphs of figure 5 can be scaled straightforward for any binary frequency and for any value of the constants in the SFR model. In order to calculate the galactic iron enrichment due to SN Ia's, these graphs can be directly implemented into a CEM without following in detail the evolution of the SN Ia binary progenitor population.

## 9. Concluding remarks

In the present paper we have calculated the temporal evolution of the rates of different types of supernovae for stellar populations with and without binaries. These computations illustrate the typical timescales on which SNe occur. Of particular interest for the study of galactic chemical evolution and the effects of SN Ia's are the timescales of the SN Ia's. To determine the rates we use a detailed population number synthesis code including single and binary stars. We then simulated the temporal evolution of the SN rates in galaxies. A major conclusion is that assuming a high interacting binary frequency, it is possible to explain the distribution and rates of supernovae in spiral galaxies in general, in our own galaxy in particular.

**Appendix A:  Stellar wind mass loss rate formalisms in massive stars**

The evolution of a massive star is critically affected by three stellar wind mass loss phases: the OB phase including the eventual luminous blue variable (LBV) phase, the red supergiant (RSG) phase and the WR phase. We treat these as follows.

*A.1. Stars with initial mass $\geq 40\ M_o$*

Based on the observations of LBVs one may suspect that stars with an initial mass > 40 $M_o$ experience an LBV phase at the end of core hydrogen burning and/or hydrogen shell burning which is accompanied by a very violent stellar wind mass loss phase. In the solar neighbourhood, the lack of RSGs with initial mass > 40 $M_o$ (corresponding roughly to stars with $M_{bol} \leq -9.5$) may be attributed to this process so that a working hypothesis for stellar evolutionary calculations may be the following:

*the $\dot{M}$ during the LBV + RSG phase of a star with initial mass larger than 40 $M_o$ must be sufficiently large to assure a RSG phase which is short enough to explain the lack of observed RSGs with $M_{bol} \leq -9.5$.*

Also the Magellanic Clouds show a deficiency of RSGs with $M_{bol} \leq -9.5$ so that the foregoing hypothesis is likely to apply for lower metallicity regions as well.

*A.2. RSG mass loss*

In our stellar evolutionary code we apply since 1997-1998 an RSG stellar wind mass loss formalism that is based on observations of Jura (1987) and Reid et al. (1990). More details are given in the papers cited in section 2. The effects of RSG mass loss on evolutionary calculations for single stars are very important for stars with an initial mass 20 $M_o$ < M < 40 $M_o$. We use stellar evolutionary calculations with a $\sqrt{Z}$-dependent $\dot{M}$-formalism for RSGs.

*A.3. WR mass loss*

Using a hydrodynamic atmosphere code where the stellar wind is assumed to be homogeneous, Hamann et al. (1995) determined $\dot{M}$–values for a large number of WR stars. Since then evidence has grown that these winds are clumpy and that an homogeneous model overestimates $\dot{M}$, typically by a factor 2-4 (Hillier, 1996; Moffat, 1996; Schmutz, 1996; Hamann and Koesterke, 1998).

In the period 1997-1998 massive single star and binary star evolutionary calculations were published (Vanbeveren et al., 1998a, b, c) for which we adopted the lower WR mass loss rates as inferred from the studies listed above. We assumed a (simple) relation between $\dot{M}(M_o/yr)$ and the stellar luminosity L (in $L_o$) i.e. $\log(-\dot{M}) = a\log L + b$, and we tried to find values for the constants a and b by accounting for the following criteria and observations known at that time:

- the WN5 star HD 50896 (WR 6) has a luminosity $\log L = 5.6$-$5.7$ and a $\log(-\dot{M}) = -4.4 \pm 0.15$ (Schmutz, 1997)
- the $\log(-\dot{M})$ of the WNE component of the binary V444 Cyg (WR 139) derived from the observed orbital period variation is ~ -5 (Khaliullin et al., 1984; Underhill et al., 1990). Its orbital mass is 9 $M_o$ and using a mass-luminosity relation holding for WNE-binary components (Vanbeveren and Packet, 1979; Langer, 1989) it follows that its $\log L = 5$
- the observed masses of black hole components in X-ray binaries indicate that stars with an initial mass > 40 $M_o$ should end their life with a mass larger than 10 $M_o$ (= the mass of the star at the end of CHeB)
- the WN/WC number ratio predicted by stellar evolution depends on the WR mass loss formalism. Therefore, last but not least, we looked for a and b values which predict the observed WN/WC number ratio (= 1) for the solar neighbourhood.

This exercise allowed us to propose the following relation:

$$\log(-\dot{M}) = \log L - 10 \qquad (A1)$$

Since then more WR stars have been investigated with detailed atmosphere codes including the effects of clumping. They are listed in table A1 and depicted in figure A1. This figure illustrates that relation (A1) still fits fairly well these new observations. Interestingly, the WC6 star OB10-WR1 in the association OB10 of M31 has been investigated by Smartt et al. (2001) and also fits relation (A1) [the star has been plotted in figure A1].

| WR number | log L | log(-$\dot{M}$) | DVB | NL | Ref |
|---|---|---|---|---|---|
| WR6 | 5.45 | -4.4 | -4.5 | -4.8 | [1] |
| WR147 | 5.65 | -4.6 | -4.3 | -4.4 | [2] |
| WR111 | 5.3 | -4.8 | -4.7 | -4.9 | [3] |
| WR90 | 5.5 | -4.6 | -4.5 | -4.7 | [4] |
| WR135 | 5.2 | -4.9 | -4.8 | -4.8 | [4] |
| WR146 | 5.7 | -4.5 | -4.3 | -4.4 | [4] |
| WR11 | 5 | -5.1 | -5 | -5 | [5] |
| WR123 | 5.7 | -4.14 | -4.3 | -4.5 | [6] |
| WR139 | 5 | -5 | -5 | -4.5 | [7] |

Table A1: The luminosity and the mass loss rates of WR stars determined with NLTE atmosphere models and assuming non-homogeneous stellar winds. We compare with values predicted by equation (A1) (DVB) and with the formula proposed by Nugis and Lamers (2000, NL). Ref. [1] = Schmutz (1997), [2] = Morris et al. (2000), [3] = Hillier and Miller (1999), [4] = Dessart et al. (2000), [5] De Marco et al. (2000), [6] = Nugis et al. (1998), [7] = Underhill et al. (1990)

The dependency on metallicity of WR mass loss rates deserves some attention. When the stellar wind is radiation driven, one expects that iron is the main driver and that the $\dot{M}$ depends mainly on the iron abundance $X_{Fe}$. Our evolutionary library contains calculations with and without a $\sqrt{X_{Fe}}$ -dependency of the $\dot{M}$ during the hydrogen deficient CHeB phase of a massive star.

*A4. Evolutionary results*

In 1997-1998 we calculated evolutionary tracks of massive single stars and binary components adopting the RSG mass loss formalism for single stars discussed above and equation (A1) for the stellar wind mass loss rate at the moment that the star becomes a hydrogen deficient core helium burning object. These tracks are part of our library. All our results published since 1997-1998 (see the papers cited section 2) rely on this library, thus rely on equation (A1). Of particular importance for the SN rates and BH formation are the pre-SN evolutionary masses (=$M_f$) predicted by our calculations. They are illustrated in figure A2. We notice that:

- Galactic massive stars have a pre-SN mass between 2 $M_o$ and 20-25 $M_o$ which allows to explain in a straightforward way the observed BH masses in the standard high mass X-ray binary Cyg X-1 and in a number of low mass X-ray binaries.
- Due to the Z-dependence of the WR (and/or RSG) stellar wind mass loss rate, the pre-SN masses of massive stars may be significantly larger in low Z-regions than in our galaxy.

**Appendix B: The period evolution during a non-conservative RLOF**

When mass leaves the binary one has to account for the loss of orbital angular momentum. In all our PNS results since 1997-1998, we used a formalism described by Soberman et al. (1997). Matter leaves a binary through the second Lagrangian point $L_2$, and settles in a circumbinary ring with radius $\eta A$ (A = distance between both components). It is straightforward then to calculate the variation of the binary period (see also Vanbeveren et al., 1998b). A "bare-minimum" for the circumbinary radius is found for $\eta$ equal 1.3, which corresponds to the distance between $L_2$ and the centre of mass of the binary. However as argued by Soberman et al. this ring is unstable and is likely to fragment, and to fall back on the binary components. The first stable ring corresponds to $\eta \approx 2.25$. In our calculations we adopt the latter value.

**References.**

Abt, H.A., 1983, ARA&A, 21, 343.

Bailyn, C.D., Jain, R.K., Coppi, P., Orosz, J.A., 1998, ApJ, 520, 696.

Belczynski, K., Kalogera, V., Bulik, T., 2002, ApJ, 572, 407.

Bloom, J.S., et al., 1999, Nature, 401, 453-456.

Brandt, W.N., Podsiadlowski, P., Sigurdssen, S., 1995, MNRAS, 277, L35.

Cappellaro, E., Evans, R., Turatto, M., 1999, A&A, 351, 459.

Cappellaro, E., Turatto, M., 2001, *in The Influence of Binaries on Stellar Population Studies*, ed. D. Vanbeveren, Kluwer Acad. Pub., Dordrecht, p.199.

Chiappini, C., Matteucci, F., Beers, T. C., Nomoto, K., 1999, AJ, 515, 226.

Chiosi, C., 1980, A&A, 83, 206.

De Donder, E., Vanbeveren, D., (paper I) 1998, A&A, 333, 557.

De Donder, E., Vanbeveren, D., 2002a, New Astronomy 7, 55.

De Donder, E., Vanbeveren, D., 2002b, New Astronomy (in press).

De Donder, E., Vanbeveren, D., Van Bever, J., 1997, A&A, 318, 812.

De Donder, E., Vanbeveren, D.: 2001, in *The Influence of Binaries on Stellar Population Studies*, ed. D. Vanbeveren, Kluwer Academic Publishers: Dordrecht, p. 535.

De Kool, M.: 1990, ApJ, 358, 189.

De Marco, O., Schmutz, W., Crowther, P. A., Hillier, D. J., Dessart, L., de Koter, A., Schweickhardt, J. : 2000, A&A, 358,187.

Fryer, C., 1999, ApJ, 522, 413.

Galama, T.J., et al., 1998, Nature, 395, 670-672.

Galama, T.J., et al., 1999, A&AS, 138, 465-466

Garmany, C.D., Conti, P.S., Massey, P., 1980, ApJ, 242, 1063.

Hachisu, I., Kato, M., Nomoto, K., 1996, ApJ, 470, L97.

Hachisu, I., Kato, M., Nomoto, K., Umeda, H., 1999a, ApJ, 519, 314.

Hachisu, I., Kato, M., Nomoto, K.: 1999b, ApJ, 522, 487.

Hamann, W.-R., Koesterke, L., 1998, A&A, 333, 251.

Hamann, W.-R., Koesterke, L., 2000, A&A, 360, 647.

Hamann, W.-R., Koesterke, L., Wessolowski, U., 1995, A&A, 299, 151.

Hillier, D.J., 1996, *in Wolf-Rayet stars in the Framework of Stellar Evolution*, eds. J.M. Vreux, A. Detal, D. Fraipont-Caro, E. Gosset, G. Rauw, Université de Liège, p. 509.

Hillier, D.J., Miller, D.L., 1998, ApJ, 496, 407.

Hogeveen, S.J., 1991, Ph.D. Thesis, University of Amsterdam.

Hogeveen, S.J., 1992, Ap&SS, 196, 299.

Iben, I., Jr., Tutukov, A., 1984, ApJSS, 54, 335.

Iwamoto, K., Brachwitz, F. et al., 1999, ApJSS, 125, 439.

Jura, M., 1987, ApJ, 313, 743.

Kahabka, P., van den Heuvel, E.P.J., ARA&A, 1997, 35, 69.

Kalogera, V., Belczynski, K.: 2001, in *The Influence of Binaries on Stellar Population Studies*, ed. D. Vanbeveren, Kluwer Academic Publishers: Dordrecht, p. 447.

Khaliullin, K.F., Khaliullin, A.I., Cherepashschuk, A.M., 1984, Soviet Astron. Letters 10, 250.

King, A.R., Pringle, J.E., Wickramasinghe, D.T., 2001, MNRAS, 320, L45.

Kobayashi, C., Tsujimoto, T., Nomoto, K., 2000, ApJ, 539, 26.

Kobayashi, C., Tsujimoto,T., Nomoto, K., Hachisu, I., Kato, M., 1998, ApJ, 503, L155.

Landau, L.D., Lifshitz, E.M., 1959, Classical Theory of Fields, Pergamon Press.

Langer, N., 1989, A&A, 210, 93.

Lorimer, D.R., Bailes, M., Harrison, P.A., 1997, MNRAS, 289, 592.

Lubow, S.H., Shu, F.H., 1975, ApJ, 198, 383.

MacFayden, A.I., Woosley, S.E., 1999, ApJ, 524, 262.

Mason, B.D., Hartkopf, W.L., Holdenried, E.R., Rafferty, T.J., 2001, AJ, 121, 3224.

Matteucci, F., Recchi, S.: 2001, ApJ, 558, 351.

Moffat, A.F.J., 1996, in *Wolf-Rayet stars in the Framework of Stellar Evolution*, eds. J.M. Vreux, A. Detal, D. Fraipont-Caro, E. Gosset, G. Rauw, Université de Liège, p. 553.

Morris, P.W., van der Hucht, K.A., Crowther, P.A., Hillier, D.J., Dessart, L., Williams, P.M., Willis, A.J., 2000, A&A, 353, 624.

Nakamura, T., Umeda, H., et al., 2001, ApJ, 555, 880.

Nelemans, G., Tauris, T.M., van den Heuvel, E.P.J., 1999, A&A, 352, L87-L90.

Neo, S., Miyaji, S., Nomoto, K., Sugimoto, D.: 1977, Publ. Astron. Soc. Japan 29, 249.

Nomoto, K., Thielemann, F.K., Yokoi, K., 1984, ApJ, 286, 644.

Nugis, T., Lamers, H.J.G.L.M., 2000, A&A, 360, 227.

Nugis,T., Crowther, P.A., Willis, A.J., A&A, 333, 956.

Paczynski, B., 1998, ApJ, 494, L45.

Popova, E.I., Tutukov, A.V., Yungelson, L.R., 1982, Astron. Space Sci., 88, 55.

Reichart, D.E., et al., 1999, APJ, 517, 692.

Reid, N., Tinney, C., Mould, J., 1990, ApJ, 348, 98.

Salpeter, E.E., 1955, ApJ, 121, 161.

Scalo, J.M., 1986, Fundam. Cosmic Phys., 11, 1.

Schaller, G., Shaerer, D., Meynet, G., Maeder, A., 1992, A&AS, 96, 269.

Schmutz, W., 1996, in *Wolf-Rayet stars in the Framework of Stellar Evolution*, eds. J.M. Vreux, A. Detal, D. Fraipont-Caro, E. Gosset, G. Rauw, Université de Liège, p. 553.

Schmutz, W., 1997, A&A, 321, 268.

Segretain, L., Chabrier, G., Mochkovitch, R., 1997, ApJ, 481, 355.

Shara, M., Hurley, J.R.: 2002, ApJ, 571, 830.

Smartt, S. J., Crowther, P. A., Dufton, P. L., Lennon, D. J., Kudritzki, R. P., Herrero, A., McCarthy, J. K., Bresolin, F. 2001, MNRAS, 325, 257

Soberman, G. E.; Phinney, E. S.; van den Heuvel, E. P. J., 1997, A&A, 327, 620.

Sparks, W.M., Stecher, T.P., 1974, ApJ, 188, 149.

Strom, R.G., 1994, A&A, 288, L1.

Talbott, R.J., R.J., Arnett, D.W., 1975, ApJ, 197, 551.

Underhill, A.B., Greve, G.R., Louth, H., 1990, PASP 102, 749.

Van Bever, J., Vanbeveren, D., 1998, A&A, 334, 21.

Van Bever, J., Vanbeveren, D., 2000, A&A, 358, 462.

Van Rensbergen, W., 2001, in *The Influence of Binaries on Stellar Population Studies*, ed. D. Vanbeveren, Kluwer Acad. Pub., Dordrecht, p.21.

Vanbeveren, D., De Donder, E., Van Bever, J., Van Rensbergen, W., De Loore, C.: 1998c, NewA 3, 443.

Vanbeveren, D., De Loore, C., 1994, A&A, 290, 129.

Vanbeveren, D., Herrero, A., Kunze, D., Van Kerkwijk, M., 1994, Space Sci. Rev., 66, 395.

Vanbeveren, D., Packet, W., 1979, A&A, 80, 242.

Vanbeveren, D., Van Bever, J., De Donder, E., 1997, A&A, 317, 487

Vanbeveren, D., Van Rensbergen, W., De Loore, C.: 1998a, The Astron. Astrophys. Review 9, 63.

Vanbeveren, D., Van Rensbergen, W., De Loore, C.: 1998b, monograph *The Brightest Binaries*, eds. Kluwer Academic Publishers: Dordrecht.

Vanbeveren, D.: 2000, in *The Evolution of the Milky Way*, eds. F. Matteucci and F. Giovanelli, Kluwer Academic Publishers: Dordrecht, p. 139.

Vanbeveren, D.: 2001, in *The Influence of Binaries on Stellar Population Studies*, ed. D. Vanbeveren, Kluwer Academic Publishers: Dordrecht, p. 249.

van den Heuvel, E.P.J., Bhattacharya, D., Nomoto, K., Rappaport, S.A., 1992, A&A, 262, 97-105.

van Paradijs, J., 1999, Science, 286, 693-695.

Webbink, R.F., 1984, ApJ, 277, 355.

Whelan, J., Iben, I. Jr., 1973, ApJ, 186, 1001.

Woosley, S. E., Weaver, T. A., 1995, ApJ.SS, 101.

Yungelson, L., Livio, M., 1998, ApJ, 497, 168.

Yungelson, L., Livio, M.: 2000, ApJ, 528, 108.

Yungelson, L.R., Nelemans, G., Portegies Zwart, S.F., Verbunt, F., 2001, , *in The Influence of Binaries on Stellar Population Studies*, ed. D. Vanbeveren, Kluwer Acad. Pub., Dordrecht, p.339.

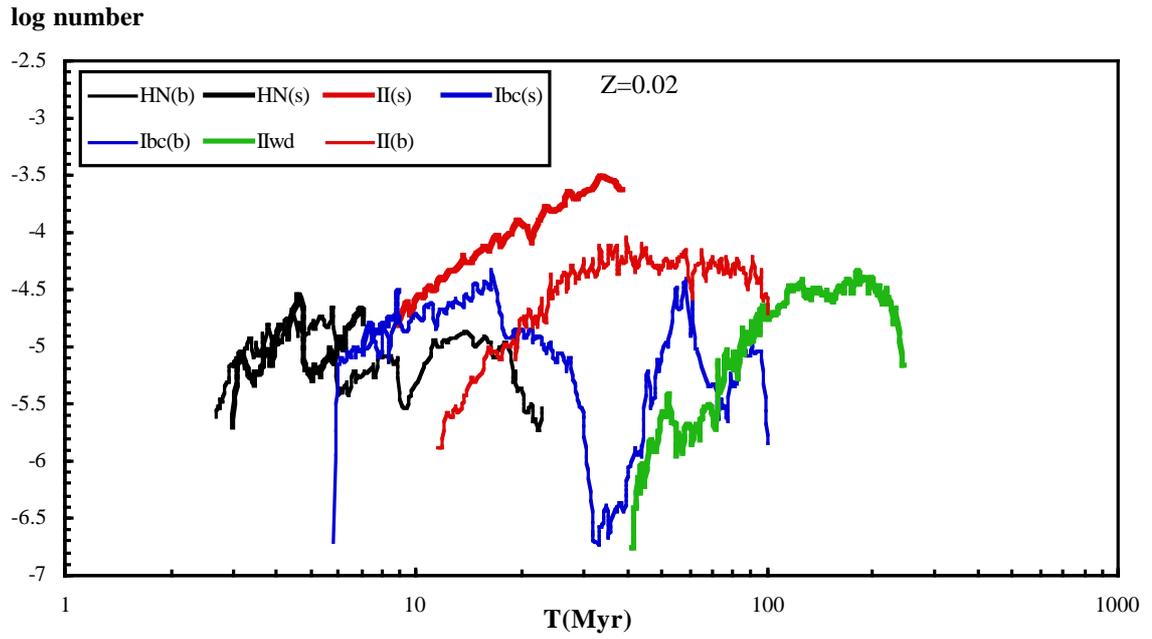

**Figure 1a:** The time evolution of the core-collapse SN rates after an instantaneous starburst with Z=0.02. The times are expressed in units of million years. We separately consider single stars (curves labelled s) and binary components (curves labelled b).

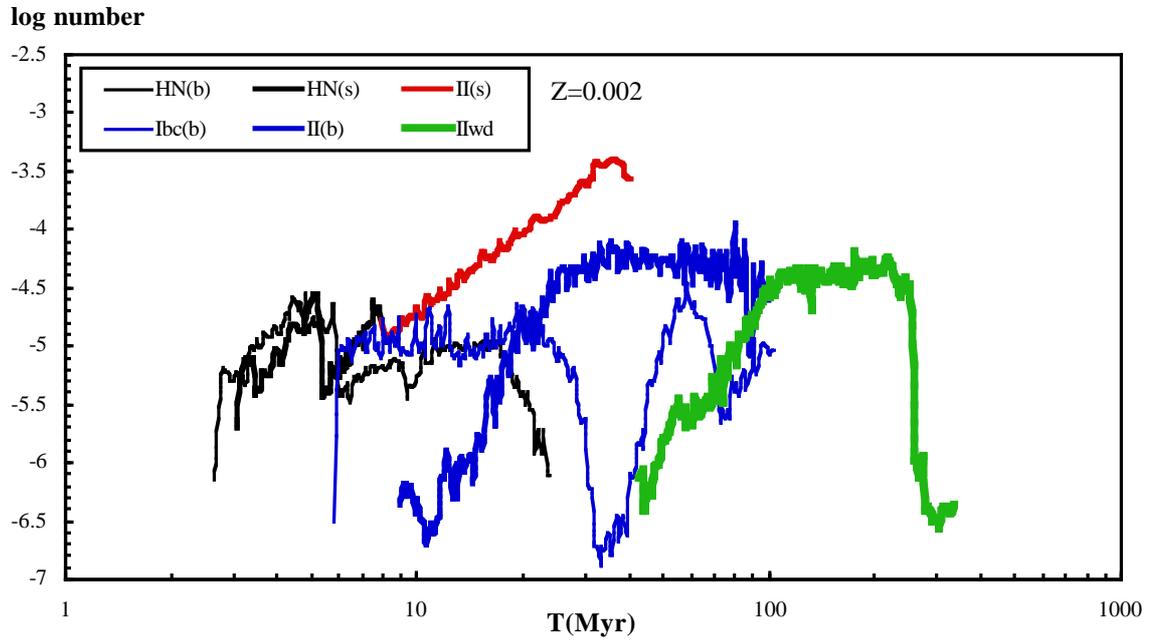

**Figure 1b:** The same as figure 1a but for Z=0.002.

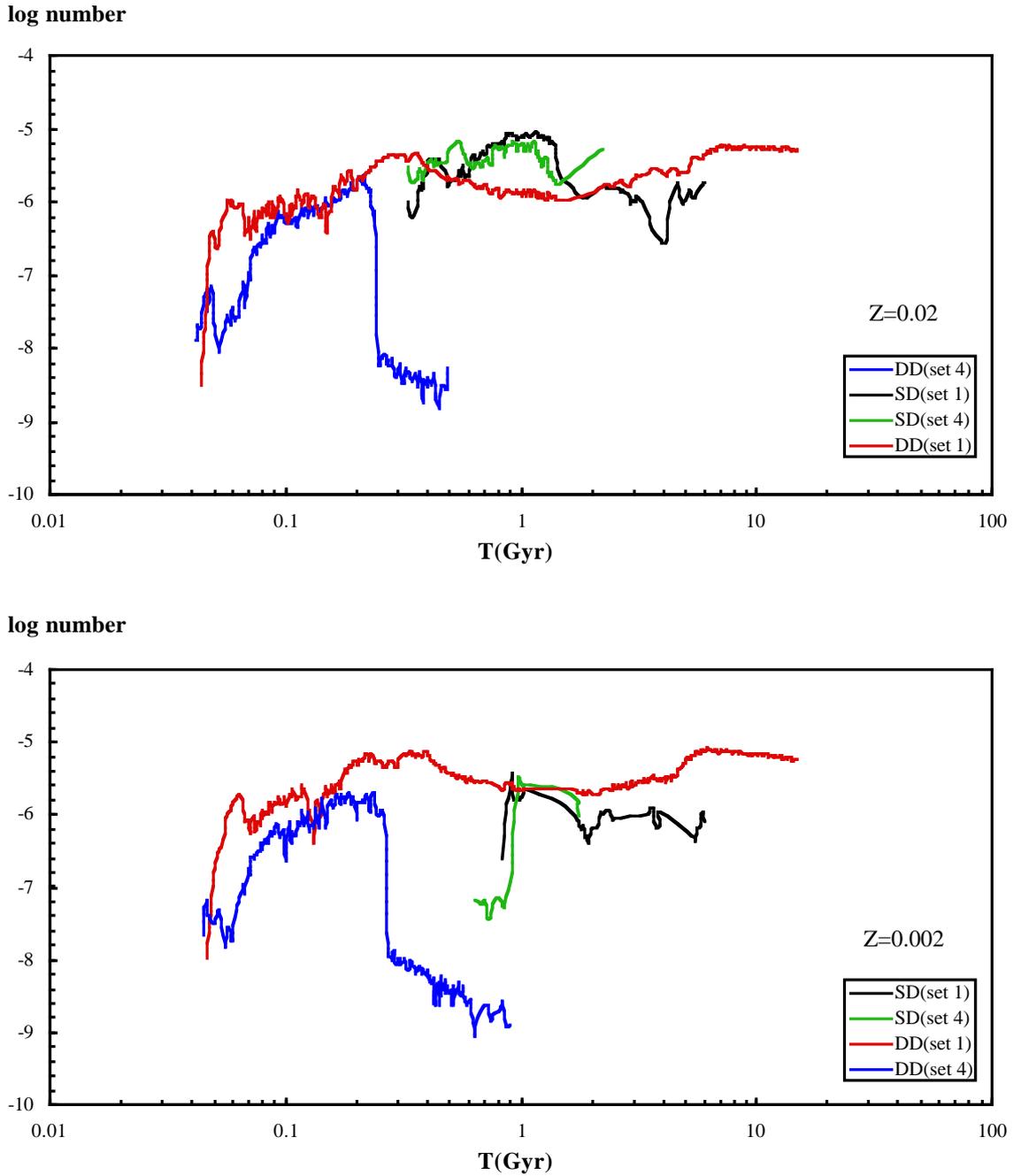

**Figure 2:** The time evolution of the SNIa rates after an instantaneous starburst for Z=0.02 (upper figure) and Z=0.002 (lower figure). The computations are made separately for the SD model and the DD model in combination with the PNS parameter sets 1 and 4, to illustrate the effect of $\beta_{max}$ and $\alpha$ which are the most affecting ones in the evolution of the SNIa progenitor binary systems.

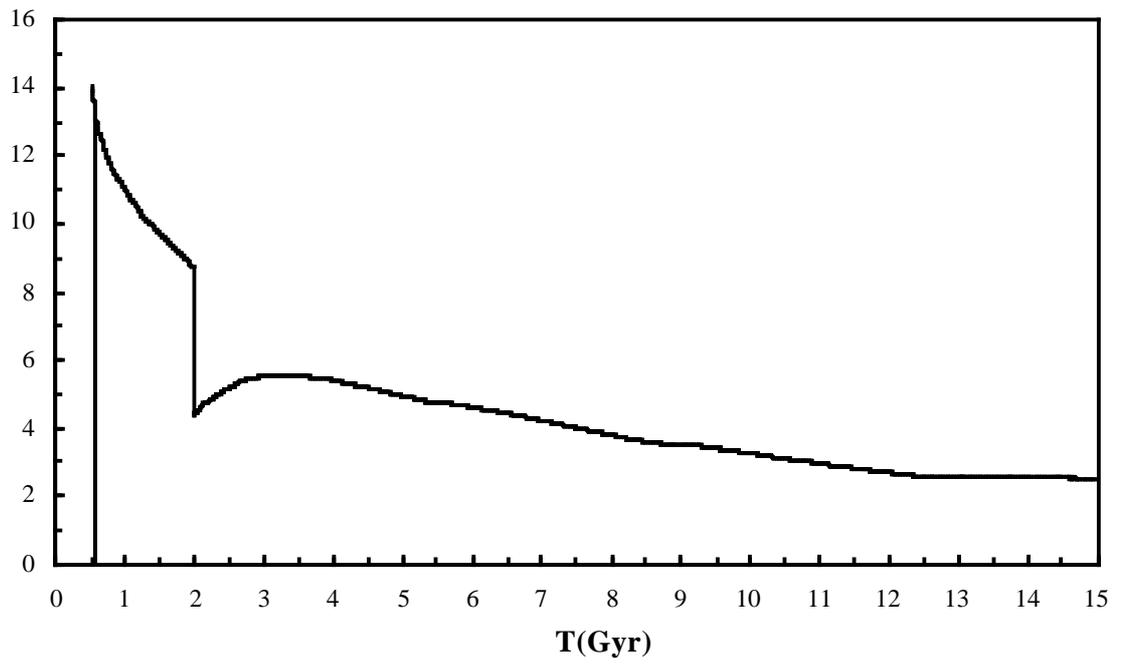

**Figure 3:** The typical star formation rate as a function of time for the two-infall galaxy formation scenario.

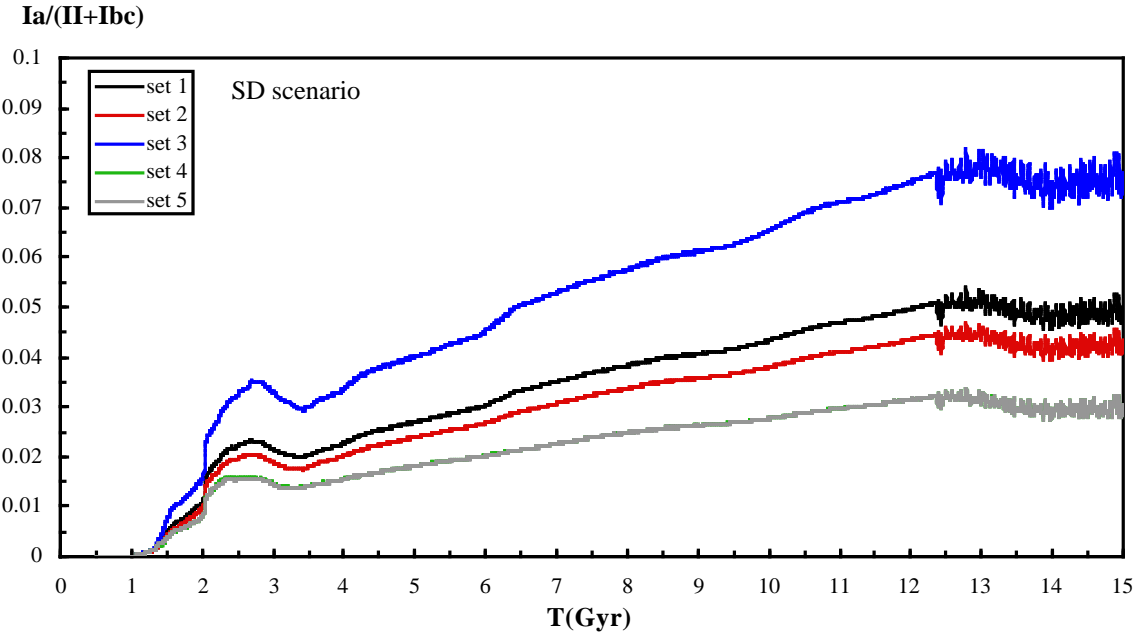

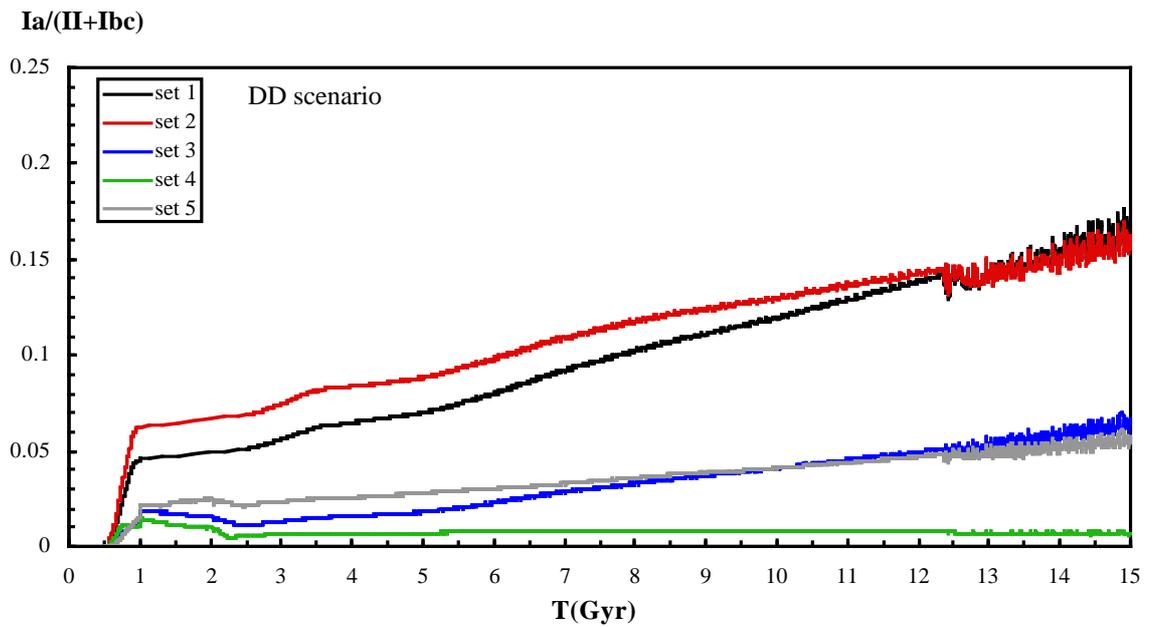

**Figure 4a:** The predicted time evolution of the SN number ratio Ia/(II+Ibc) in a spiral galaxy computed for the SD (upper figure) and the DD (lower figure) scenario. The adopted binary frequency is 40%.

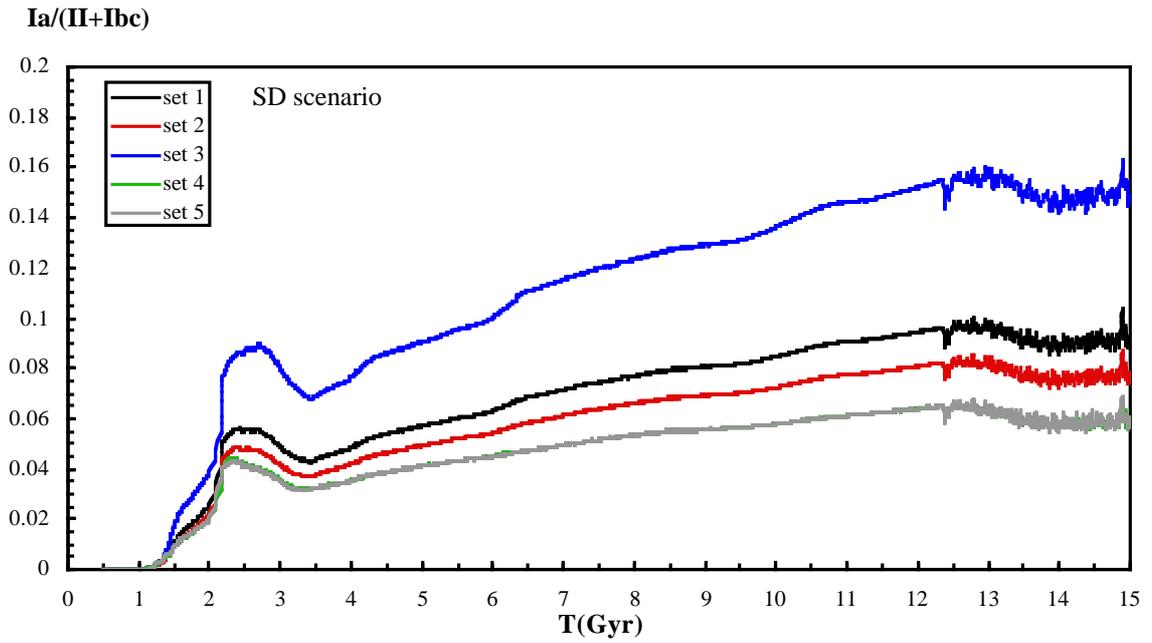

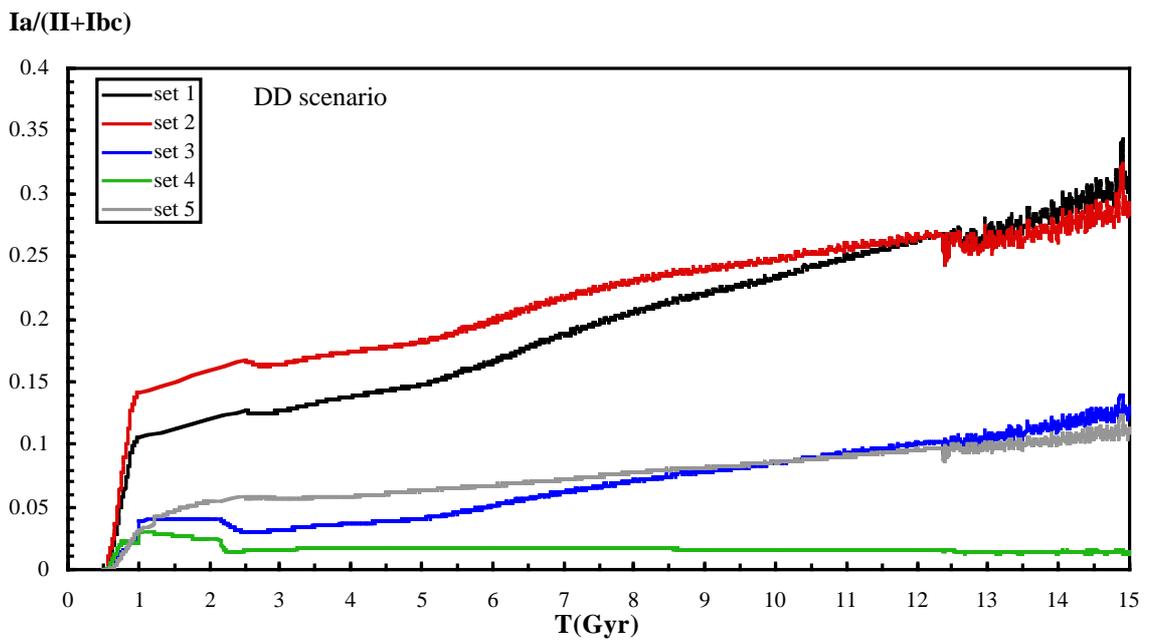

**Figure 4b:** The same as figure 4a but for a constant binary frequency of 70%.

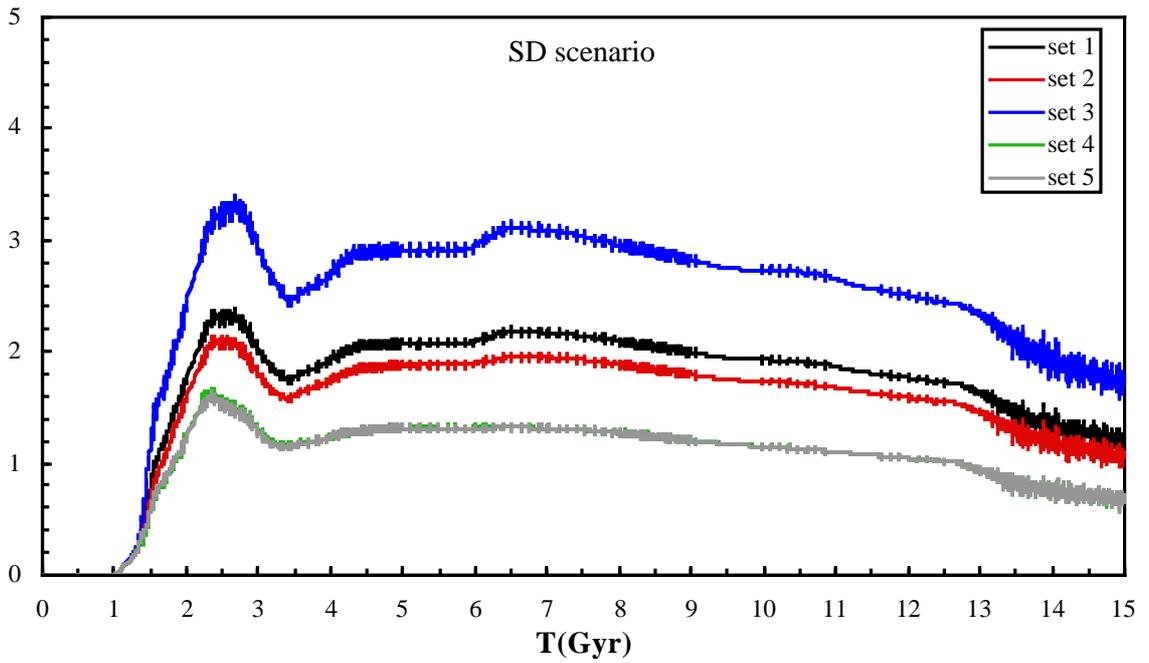

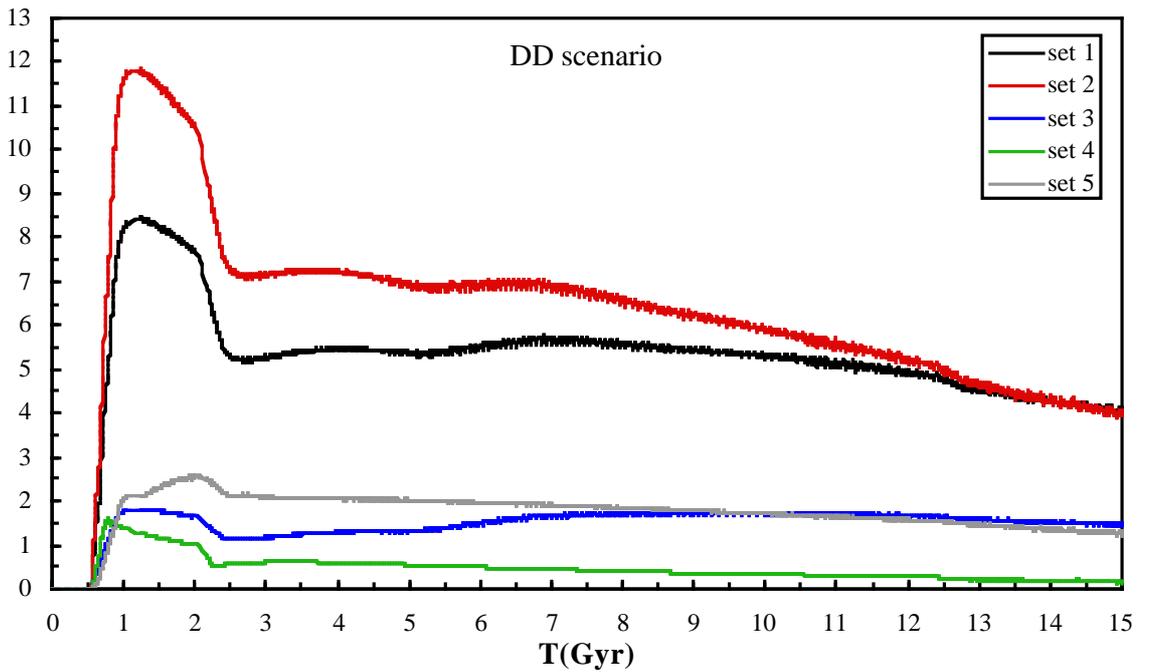

**Figure 5:** The predicted time evolution of the SN Ia rates ( per millennium) in a spiral galaxy separately computed for the SD (upper figure) and DD (lower figure) scenario. The computations are made for a constant binary frequency of 70%.

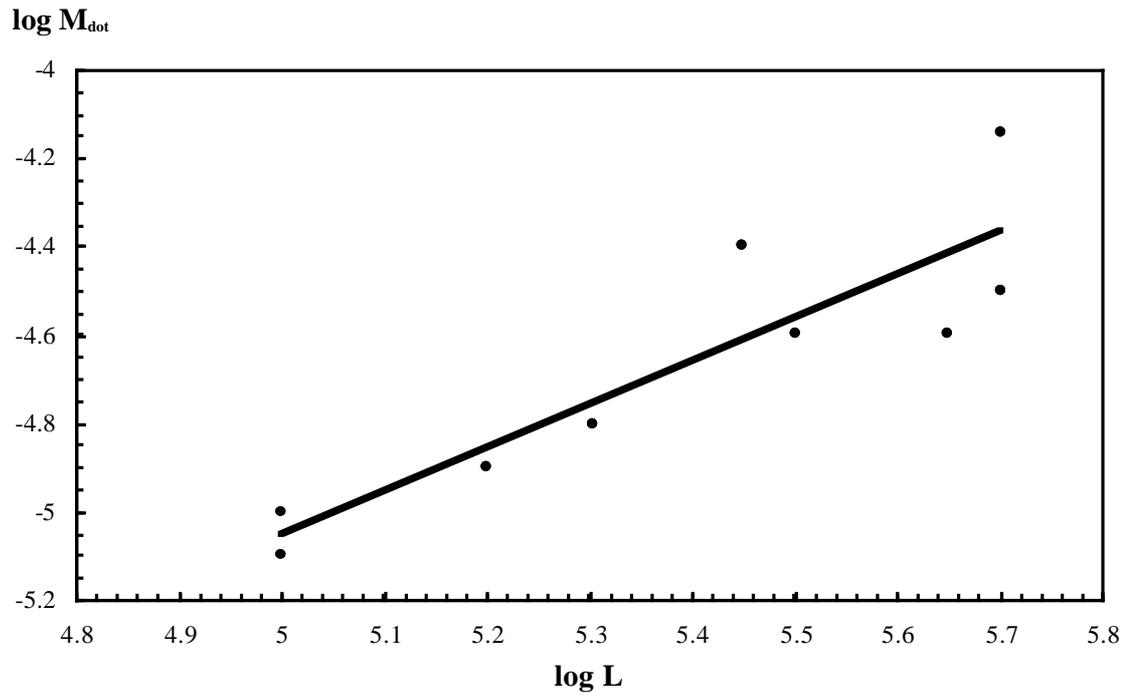

**Figure A1:** The observed mass loss rates versus log L for WR stars. The trendline is given by equation A1 adopted here.

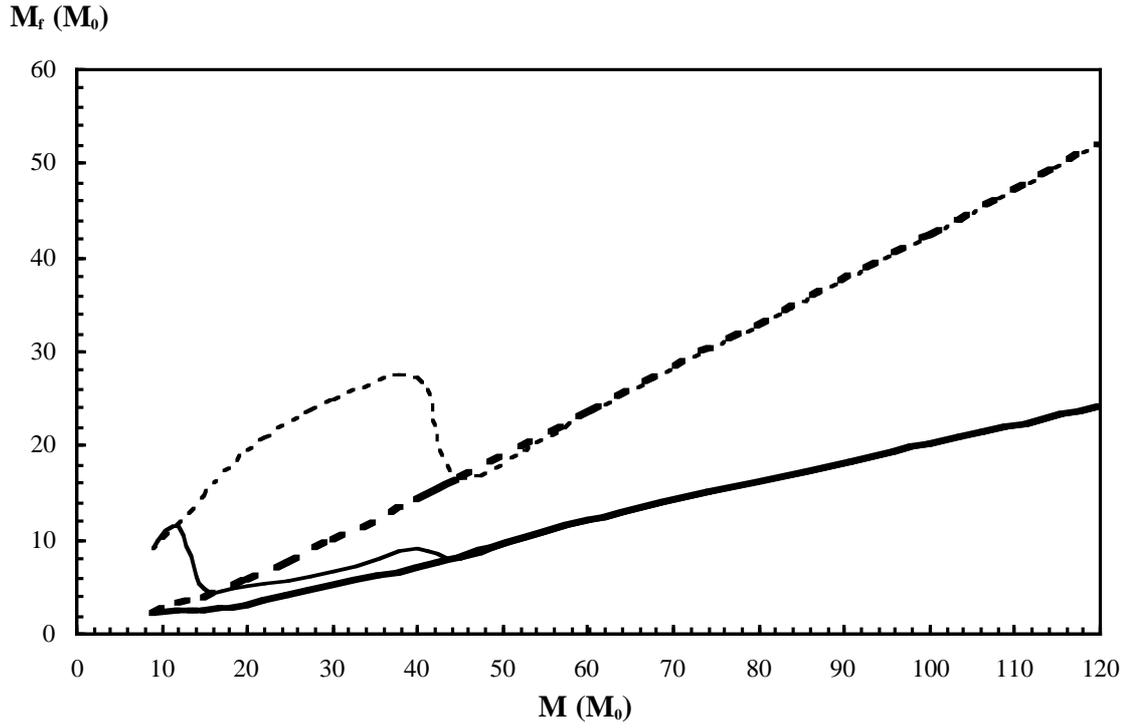

**Figure A2:** The pre-SN masses ($M_f$) as a function of the initial mass M (on the ZAMS) of massive single (thin lines) and interacting primary (thick lines) stars as predicted by our calculations. The computations are made for Z=0.02 (full lines) and Z=0.002 (dashed lines).